\documentclass[superscriptaddress,aps, prd, reprint, nofootinbib]{revtex4-2}
\usepackage{amssymb}
\usepackage{orcidlink}
\usepackage{amsmath}
\usepackage{mathtools}
\usepackage{siunitx}
\usepackage{aas_macros}
\newcommand{\calE}{\mathcal{E}}
\newcommand{\revision}[1]{{#1}} 
\newcommand{\lab}[1]{{\mathrm{#1}}}
\DeclareSIUnit\solarmass{\ensuremath{\mathrm{M}_\odot}}
\DeclareSIUnit\parsec{pc}
\DeclareSIUnit\year{yr}
\usepackage{hyperref}
\hypersetup{
colorlinks=true, % false: boxed links; true: colored links
linkcolor=blue, % color of internal links
citecolor=blue, % color of links to bibliography
filecolor=blue, % color of file links
urlcolor=blue, % color of external links
runcolor=blue
}

\usepackage{amssymb}

\begin{document}

\title{Dark matter, black holes, and gravitational waves}

\author{Gianfranco Bertone\,\orcidlink{0000-0002-6191-1487}\,}

\affiliation{Gravitation Astroparticle Physics Amsterdam (GRAPPA),
University of Amsterdam, Amsterdam, 1098 XH, Netherlands}

\begin{abstract}
%% Text of abstract
The formation and growth of black holes can strongly influence the distribution of dark matter around them. I discuss here the different types of dark matter overdensities around black holes, including dark matter cusps, spikes, mounds, crests, and gravitational atoms. I then review recent results on the evolution of a black holes binary in presence of dark matter, focusing on the energy transfer between binary and dark matter induced by dynamical friction. Finally, I present the prospects for studying dark matter with gravitational wave observations, and argue that future interferometers might be able to detect and characterise dark matter overdensities around black holes.
\end{abstract}
\maketitle

\section{Introduction}

In this contribution to the proceedings of the Nobel Symposium on dark matter, I review a number of new and exciting results at the interface between dark matter, black holes, and gravitational waves, and I discuss the future prospects and potential directions for this emerging field of research. The focus of this contribution may seem surprising at first sight, as the evidence for dark matter historically arose from observations in the extremely weak gravitational field regime \cite{2018RvMP...90d5002B,2005PhR...405..279B}, while black holes and gravitational waves are typical strong gravity phenomena. 

Yet, several independent lines of research reveal many interesting points of convergence \cite{2019arXiv190710610B}. We know for instance since the 1990s that dark matter {\it cusps} are predicted to form at the centers of galactic halos \cite{1996ApJ...462..563N}. And we have direct and indirect evidence that massive black holes reside at the center of most galaxies (see e.g. Ref. \cite{2021NatRP...3..732V} for a recent review). We also know that the presence and evolution of black holes  inevitably influence the surrounding matter distribution, which implies that dark matter can in principle form very high density regions around black holes and alter their phenomenology (see the discussion and references below). 

Large dark matter overdensities may also form around black holes via other channels. They might be seeded by primordial black holes, as dark matter decouples from the Hubble flow (see e.g. Refs. \cite{1985ApJS...58...39B,2007ApJ...665.1277M,2007ApJ...662...53R,2021JCAP...08..053B} and references therein), or they may grow via the process of superradiance, if the mass of the dark matter particles is so light that its Compton wavelength is comparable with the size of the black hole (see e.g. Refs.  \cite{1971JETPL..14..180Z,1972JETP...35.1085Z,1973JETP...37...28S,1974JETP...38....1S,2017PhRvD..96b4004E,2017PhRvL.119d1101E}, as well as Ref. \cite{2015LNP...906.....B} for a recent review).  Whatever the physical process responsible for the formation of dark matter overdensities around black holes, if dark matter is made of self-annihilating particles, black holes can effectively act as {\it annihilation boosters}. If that's not the case, dark matter can reach extremely high densities and induce “environmental” effects that can be probed with GW experiments \cite{2019arXiv190710610B}.
 
The paper is organised as follows: in Sec.~\ref{sec:spikes} I review the different types of dark matter overdensities around black holes; in Sec.~\ref{sec:dynamics} I describe the evolution of a black holes binary in presence of dark matter; in Sec.~\ref{sec:gw} I comment on the prospects for characterising dark matter with gravitational wave observations. 
  
%% main text
\section{Matter around Black Holes}
\label{sec:spikes}

\textbf{Collisional stellar cusps.} The study of the dynamics of matter around black holes has a long history. In the 1970s and 1980s many authors studied the evolution of stellar clusters around a massive black hole, see e.g. Ref. \cite{1985IAUS..113..373S} and references therein. 
 In these systems, the two-body relaxation time is much shorter than the age of the system, the stars reach a steady-state solution resulting in a density cusp with density $\rho(r) \sim r^{-7/4}$ as shown by Bahcall and Wolf in 1976 \cite{1976ApJ...209..214B}. Their theoretical framework provides a good description of the observed stellar cusp at the center of the Milky Way, once the effects of mass segregation and continued star formation are taken into account, as discussed e.g. in Refs. \cite{2018A&A...609A..28B,2019MNRAS.484.3279P}. This has important consequences for the estimates of the rate of extreme mass ratio inspirals discussed below \cite{2011CQGra..28i4017A,2014MNRAS.437.1259B,2019MNRAS.483..152A}.

\textbf{Adiabatic compression of isothermal systems.} If the relaxation time is long compared with the age of the system, then the distribution of stars (or of dark matter particles) depends on the formation history of the system. A particular case that has been explored in great detail is the `adiabatic growth' scenario, in which the central black hole grows on a timescale that is much longer than the dynamical timescale. This problem was considered by Peebles in the framework of a spherical and isothermal stellar cusp surrounding a massive black hole in an elliptical galaxy \cite{1972GReGr...3...63P}. Peebles showed that the conservation of the adiabatic invariants (in this case the radial action and angular momentum), led to a density cusp $\rho(r)\sim r^{-3/2}$. 
The result was later confirmed by Young in 1980 with an iterative numerical approach, always under the assumption of an isothermal distribution of stars \cite{1980ApJ...242.1232Y}. 

\textbf{Adiabatic compression of power-law profiles.} The same  numerical approach was then extended by Quinlan, Hernquist and Sigurdsson \cite{1995ApJ...440..554Q} to the case of cusps of collisionless matter described by the $\gamma$-models introduced by Dehnen \cite{1993MNRAS.265..250D}. In this case, the density and the distribution function $f(E)$ diverged as power laws at the center of the distribution:
\begin{align}
    & \rho(r) \sim r^{-\gamma} \\
    & f(E) \sim [E-\phi(0)]^{-n} \, , 
\end{align}
where $E$ is the energy per unit mass and $\phi(0)$ is the gravitational potential at the center.
The conservation of the radial action implies that the final density profile after adiabatic growth is again a power-law \cite{1995ApJ...440..554Q}:
\begin{align}
   & \rho_{\rm{final}}(r) \sim r^{-\gamma_{\rm{sp}}} \\ & \gamma_{\rm{sp}}= \frac{3}{2} + n \left( \frac{2-\gamma}{4-\gamma} \right) \, .
\end{align}
For the $\gamma$ models above, $n=(6-\gamma)/(4-2\gamma)$ \cite{1993MNRAS.265..250D}, and the power-law index after adiabatic growth of the black hole is \cite{1995ApJ...440..554Q}:
\begin{equation}
\gamma_{\rm{sp}}=\frac{9-2\gamma}{4-\gamma}
\end{equation}
\textbf{Dark matter spikes.} Gondolo and Silk \cite{1999PhRvL..83.1719G} applied the results obtained by Quinlan, Hernquist, and Sigurdsson \cite{1995ApJ...440..554Q} to the dark matter cusps that several groups of authors had put forward in the 1990s \cite{1996ApJ...462..563N,1998ApJ...499L...5M,1996MNRAS.278..488Z}. They focused in particular on the case of annihilating dark matter particles, and showed that since the annihilation rate increases as $\rho^2(r)$, the existence of these adiabatically compressed profiles - dubbed \textit{spikes} - would greatly enhance the predicted annihilation flux of photons and neutrinos \cite{1999PhRvL..83.1719G}. Numerical analyses then showed that spikes extend over a region of size $r_{\rm{sp}} \approx 0.2 r_{\rm h}$, where $r_{\rm h}$ is the radius of gravitational influence of the central black hole defined from
\begin{equation}
    \int_{r_0}^{r_{\rm h}} \rho(r) 4\pi r^2 \,\mathrm{d}r = 2 M_{\rm{BH}} \, ,
\end{equation}

\textbf{Relativistic spikes.} Sadeghian, Ferrer, and Will extended the adiabatic growth formalism to fully relativistic treatment, assuming a  Schwarzschild metric for the central black hole \cite{Sadeghian:2013laa}. In this case, the dark matter density vanishes at $r=2 R_{\rm{S}}$, where $R_{\rm{S}}$ is the Schwarzschild radius, not $4R_{\rm{S}}$ as in the non-relativistic case, reaching higher values close to the black hole. The formalism was subsequently further extended to the case of a Kerr black hole, and it was shown that the black hole spin further increases  the dark matter density close to the black
hole \cite{Ferrer_2017}. 

\textbf{Formation and survival of dark matter overdensities.} The dark matter profile around a black hole  in general does not converge to a universal steady-state solution. As we saw above, it will depend on the initial phase space of the dark matter system at the formation of the black hole seed, and on the hierarchy between the various timescales of the problem -- i.e. black hole growth rate, dynamical timescale, relaxation timescale (in the case of self-interacting dark matter). The implications of non-adiabatic and off-center growth of the black hole is in particular discussed by Ullio, Zhao, and Kamionkowski \cite{2001PhRvD..64d3504U}. The problem is further complicated by the fact that massive black holes typically live in very dense environments. They may in principle undergo major mergers, which would lead to a substantial suppression of the dark matter density \cite{Merritt:2002vj,2018PhRvD..98b3536K}. 

{\bf Dark matter \textit{Mounds}.} Bertone et al. recently showed spikes are a poor approximation of the dark matter profiles for gravitational wave searches \cite{Bertone:2024wbn}. Extrapolating the power-law density profiles to small radii effectively corresponds to assuming that the central object has formed from an unrealistically small seed, whereas they typically form from the collapse of massive stars with a radius $R_{\rm{*}}$ much larger than the size of the black hole $R_{\rm{*}} \gg R_{\rm{S}}$. This assumption is harmless in the case of self-annihilating dark matter, since annihilations erase any small-scale feature imprinted by the black hole formation history on dark matter. But if dark matter is made of generic non-annihilating cold dark matter particles, it is necessary to follow the evolution of the dark matter distribution through the formation of the supermassive star that precedes the formation of the black hole, and its subsequent collapse. In this case, the dark matter density profiles are shallower than DM spikes, and can be generically referred to as DM \textit{mounds} \cite{Bertone:2024wbn}. 

\textbf{Dark matter \textit{Crests}.} Whatever the initial conditions and black hole formation history are at the center of galaxies like the Milky Way, dark matter will inevitably be scattered off stars in the stellar cusps, formed precisely through the same process. The co-evolution of the stellar and dark matter cusp would initially reduce the dark matter density in a spike \cite{2004PhRvL..92t1304M,Bertone:2005hw}. Interestingly, even if some violent process, such as an equal mass binary black hole merger, dramatically reduces the dark matter density, stars-dark matter gravitational scattering would lead to a formation of a \textit{crest} (collisionally regenerated dark-matter structure) with profile $\rho \sim r^{-3/2}$. Merritt, Harfst, and Bertone showed that, unlike spikes, the existence of crests can be robustly predicted in any nucleus old enough to have generated a Bahcall-Wolf cusp in the stars \cite{2007PhRvD..75d3517M}. 

\begin{figure*}[t]
    \centering
    \includegraphics[width=0.8\textwidth]{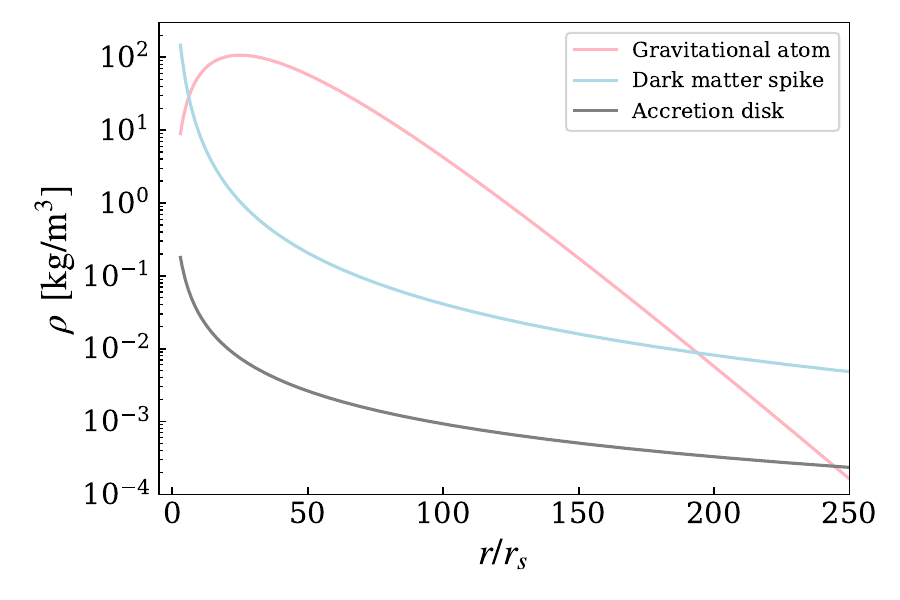}
    \caption{Initial density profiles of an accretion disc, a gravitational atom, and a dark matter spike, surrounding a $10^5\,{\rm M_\odot}$ black hole. See Ref. \cite{2023NatAs...7..943C}, from which the figure is taken, for more details.}
    \label{fig:profile}
\end{figure*}

\textbf{Dark matter \textit{mini-spikes}}. Bertone, Zentner and Silk \cite{Bertone:2005xz}, and Zhao and Silk \cite{2005PhRvL..95a1301Z} suggested that intermediate mass black holes would be more likely to carry unperturbed spikes. Ref. \cite{Bertone:2005xz} in particular estimated the number of detectable black holes in two different IMBH formation scenarios: one in which IMBHs form in rare, overdense regions at high redshift, $z\sim 20$, as remnants of Population III stars, and one in which more massive IMBHs form via direct collapse. The former scenario implied a $\rho \sim r^{-3/2}$ overdensity, arising from adiabatic growth from an initially uniform dark matter density. The latter scenario, with a steeper $\rho \sim r^{-7/3}$ profile, has been adopted as a fiducial model for many indirect dark matter detection analyses, and for the gravitational wave studies discussed below. 

\textbf{Dark matter around Primordial Black Holes.}  If primordial black holes (PBHs) exist and they do not constitute all of the dark matter, then they should carry a dark matter `dress' around them. For an in-depth discussion of the determination of the dark matter profile with analytical and numerical methods, see Ref. \cite{2021JCAP...08..053B} and references therein. Lacki and Beacom argued that if dark matter is made of WIMPs, then gamma-rays produced by the annihilation of dark matter particles around primordial black holes would exceed the gamma-ray extragalactic background unless $\Omega_{PBH}<10^{-4}$ \cite{2010ApJ...720L..67L} (see also \cite{2019PhRvD.100b3506A}). Bertone et al. showed that the successful detection of one or more PBHs by radio searches with the Square Kilometer Array, and gravitational waves searches with LIGO/Virgo and the upcoming Einstein Telescope, would set extraordinarily stringent constraints on any weak-scale extension of the Standard Model with stable relics. Finding PBHs, even in the case where they contribute a small fraction of the dark matter, would essentially rule out almost the entire parameter space of popular theories such as the minimal supersymmetric standard model \cite{2019PhRvD.100l3013B}.

\textbf{Annihilation boosters.} Just like in the case of primordial black holes, dark matter overdensities around astrophysical black holes would enormously increase the annihilation rate if dark matter is made of WIMPs. Black holes effectively act as dark matter annihilation boosters, leading to large flux of gamma-rays, anti-matter, and neutrinos, as discussed by many authors \cite{1999PhRvL..83.1719G,2001PhRvD..64d3504U,2004PhRvL..92t1304M,2007PhRvD..75d3517M,1998APh.....9..137B,2002PhRvL..88s1301M,2005PhRvD..72j3517B,2005PhRvD..72j3502B,2002MNRAS.337...98B,2006PhRvD..73f3510B,2014PhRvL.113o1302F,2004JCAP...05..007A,2004PhRvD..70k3007H,2018PhRvD..98f3527D}. 
Annihilations set an 
upper limit to the dark matter density
\begin{equation}
\rho_{\rm sp}(r_{\rm  lim})\approx m_\chi/\sigma v \,(t-t_f) \equiv \rho_{\rm lim}
\,\,\, ,
\label{eq:rholim}
\end{equation}
where $\sigma v$ is the annihilation cross section, $m_\chi$ is the dark particle mass, so $\rho(r) \sim \rho_{\rm lim}$ for $r \leq r_{\rm lim}$ (for a more detailed treatment of the so-called annihilation plateau, see \cite{2016PhRvD..93l3510S,2007PhRvD..76j3532V}).
The flux of secondary particles of energy $E$ produced by the annihilation of dark matter particles a mini-spike at distance $D$ can be expressed, assuming a Navarro-Frenk-White profile \cite{1996ApJ...462..563N} with 
$\gamma=1$ ($\gamma_{\rm sp} = 7/3$), as \cite{Bertone:2005xz}
\begin{eqnarray}
\nonumber
\Phi (E,D) & = & \Phi_0 \frac{{\rm d}N}{{\rm d}E} 
\left( \frac{\sigma v}{10^{-26} {\rm cm}^3/{\rm s}} \right)
\left( \frac{m_\chi}{100 {\rm GeV}} \right)^{-2} 
\\ 
 & \times & \left( \frac{D}{{\rm kpc}} \right)^{-2} 
\left( \frac{\rho(r_{\rm sp})}{10^2 {\rm GeV}{\rm cm}^{-3}} \right)^{2} 
\nonumber \\
 & \times & 
\left( \frac{r_{\rm sp}}{{\rm pc}} \right)^\frac{14}{3}
\left( \frac{r_{\rm lim}}{10^{-3}{\rm pc}} \right)^{-\frac{5}{3}}\, ,
\label{eq:flux}
\end{eqnarray}
where $r_{\rm sp}$ is the radius of the spike, $r_{\rm lim}$ is the radius at which $\rho=\rho_{\rm  lim}$, and $\Phi_0 = 9 \times 10^{-10} {\rm cm}^{-2}{\rm s}^{-1}$.  Interestingly, while annihilation fluxes typically scale with $\sigma v /m_\chi^2$, in this case, and every time the dark matter density saturates the maximum value in Eq.  \ref{eq:rholim}, the  DM profile itself depends on $m_\chi$ and $\sigma v$. The final luminosity of the objects is thus 
proportional to $\sim (\sigma v)^{2/7} m_\chi^{-9/7}$\cite{Bertone:2005xz}.
If spikes or mini-spikes exist and dark matter is made of WIMPs, it should be relatively easy to detect them. The fact that we haven't found bright sources of gamma-rays, anti-matter or neutrinos, means that either large dark matter overdensities do not form or that dark matter is not made of WIMPs (for a more precise assessment, see e.g. Refs. \cite{2009NJPh...11j5016B,2008PhRvD..78g2008A,2006PhRvD..73j3519B,2022JCAP...08..065F} and references therein). Spikes, mini-spikes, crests, and in general any overdensity of non-annihilating dark matter, may be probed with gravitational waves, as discussed below.

\textbf{Ultralight dark matter around black holes.} A completely different physical process may lead to the formation of large dark matter overdensities around black holes. If dark matter is made of ultralight bosons, with a Compton wavelength comparable with the size of the gravitational radius of the black hole, a rotating black hole can spontaneously shed mass and angular momentum to form a large boson cloud, via a process called superradiance \cite{1971JETPL..14..180Z,1972JETP...35.1085Z,1973JETP...37...28S,1974JETP...38....1S,2017PhRvD..96b4004E,2017PhRvL.119d1101E}. 
The structure of the boson cloud is nearly identical to that a hydrogen atom, and the black hole-cloud system is  called a \textit{gravitational atom}. The ultralight boson cloud states are $|n\ell m\rangle$ eigenstates, with wavefunction
\begin{equation}
    \psi(t,\vec r)=R_{n\ell}(r)Y_{\ell m}(\theta,\phi)e^{-i(\omega_{n\ell m}-\mu) t},
\end{equation}
where $Y_{\ell m}$ are spherical harmonics and $R_{n \ell}(r)$ are the hydrogenic radial functions, and $\mu$ is the mass of the scalar field \cite{Baumann:2019ztm,Baumann:2021fkf}. The mass density is defined as
\begin{equation}
    \rho(\vec r)=M_{c}|\psi(\vec r)|^2,
\end{equation}
where $M_{c}$ is the total mass of the cloud, which can reach a value of about $10\%$ of the central BH mass. Gravitational atoms are a subject of intense research and we cannot possibly do justice here to all the exciting literature on the subject. We describe below the main implications for gravitational wave searches, and we refer the interested reader to the review of Brito, Cardoso and Pani \cite{2015LNP...906.....B} for further details.

\section{Binary BH dynamics in presence of dark matter}
\label{sec:dynamics}

The presence of dark matter modifies the evolution of a binary black hole system in a number of ways. For the systems discussed above, the leading effect is a transfer of energy from the binary to dark matter, due to \textit{dynamical friction} ~\cite{Chandrasekhar1943a,Chandrasekhar1943b,Chandrasekhar1943c}. Eda et al. showed that the energy losses due to dynamical friction would lead to a large dephasing of the gravitational waveform of extreme mass ratio inspirals in presence of a spike of collisionless dark matter particles~\cite{2013PhRvL.110v1101E,2015PhRvD..91d4045E}. 

Kavanagh, Nichols, Bertone, and Gaggero  showed that in most realistic cases, the work done by dynamical friction is comparable to, or larger than, the total binding energy of the dark matter spike. Previous calculations of the dephasing induced by dark matter, which assumed a fixed DM density profile, were found to violate energy conservation and to overestimate the dephasing. They thus devised a new method to jointly evolve the distribution of DM and the the binary \cite{2020PhRvD.102h3006K}.
\begin{figure*}[t]
    \centering
    \includegraphics[width=0.8\textwidth]{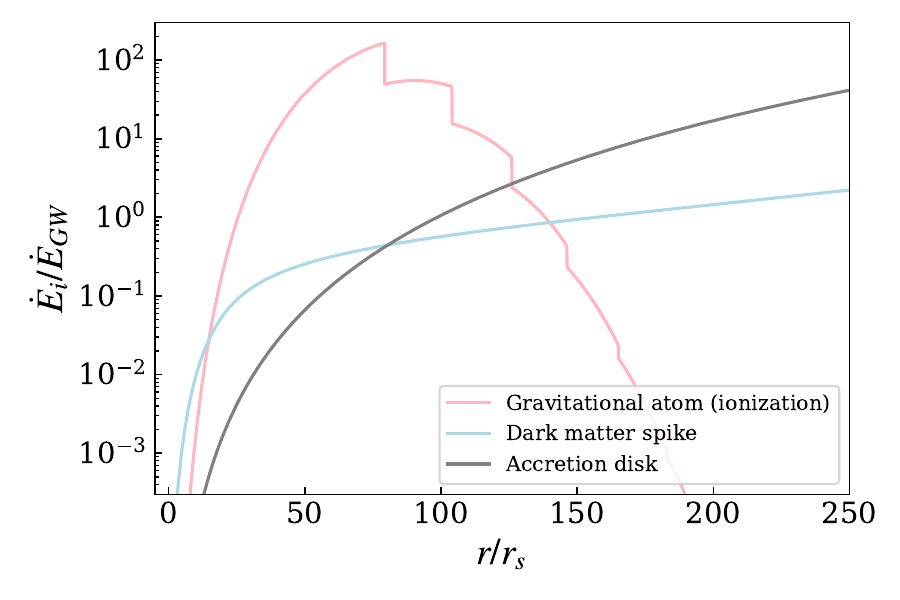}
    \caption{Energy losses induced by the three environments in fig. 1, normalised by gravitational wave energy losses. See Ref. \cite{2023NatAs...7..943C}, from which the figure is taken, for more details.}
    \label{fig:profile2}
\end{figure*}

\textbf{Energy losses for collisionless dark matter.} The evolution of the binary depends on the energy losses of the system $\dot{E} = -\dot{E}_{\rm GW}-\dot{E}_{\rm env}$.
The relative importance of the dark matter-induced energy losses $\dot{E}_{\rm env}$ with respect to the energy radiated away due to GWs $\dot{E}_{\rm GW}$ is shown in the right panel of Fig. \ref{fig:profile}, as a function of the separation of the binary, for the benchmark models in \cite{2023NatAs...7..943C}. In the case of a collisionless dark matter spike, and binaries with intermediate or extreme mass ratio, the leading environmental energy losses are due to dynamical friction. The evolution of the binary separation reads in this case \cite{2020PhRvD.102h3006K} \begin{equation}
\begin{split}
    &\dot{r} = - \frac{64\, G_N^3\, M \, m_1\, m_2}{5\, c^5\, (r_2)^3} \\ &- \frac{8 \pi\, G_N^{1/2}\, m_2 \, \log\Lambda \, r^{5/2} \, \rho(r,t) \,\xi(r, t)}{\sqrt{M} m_1 }  \, ,
    \label{eq:r_eom}
\end{split}    
\end{equation}
where $m_1$ and $m_2$ are the masses of the central and companion black holes, $M=m_1+m_2$ is the total mass of the binary, $\xi$ denotes the fraction of DM particles moving more slowly than the orbital speed, and  $\log\Lambda$ is the Coulomb logarithm. In order to evolve the time-dependent density profile $\rho(r,t)$, Kavanagh et al. started from the phase space distribution  $f$, which in a spherical system is only a function of the relative energy per unit mass and time: $f = f(\calE\revision{, t})$ \cite{2020PhRvD.102h3006K}:   
\begin{align}
\label{eq:dfdt}
    \begin{split}
        &T_\mathrm{orb} \frac{\partial f(\calE, t)}{\partial t} = - p_\mathcal{E}f(\mathcal{E}, t)  \,+ \\
        &\int \left(\frac{\calE}{\calE - \Delta\calE}\right)^{5/2} f(\calE - \Delta\calE, t)  P_{\calE-\Delta\calE}( \Delta\calE)\,\mathrm{d}\Delta\calE\,,
    \end{split}
\end{align}
where $T_\mathrm{orb} = 2\pi\sqrt{(r_2)^3/(G_N M)}$ is the orbital period, and $p_\mathcal{E} = \int P_{\mathcal{E}}(\Delta\mathcal{E}) \mathrm{d}{\Delta\mathcal{E}}$ is the total probability for a particle of energy $\mathcal{E}$ to scatter with the compact object during one orbit, obtained by integrating the probability $P_\mathcal{E}(\Delta\mathcal{E})$ that a particle with energy $\mathcal{E}$ scatters with the compact object and gains an energy $\Delta\mathcal{E}$. The time dependent dark matter density can then be recovered by integrating over velocity the phase space distribution. The evolution of the full system can thus be obtained by jointly evolving eqs. \ref{eq:r_eom} and \ref{eq:dfdt}  \cite{2020PhRvD.102h3006K,Kavanagh:2024lgq,Karydas:2024fcn}.

\textbf{Energy losses for gravitational atoms.} Baumann, Chia, Porto, and Stout studied the gravitational-wave  signatures of gravitational atoms, showing the existence of resonant transitions between bound states of the boson cloud \cite{Baumann:2019ztm,2019PhRvD..99d4001B}. Baumann, Bertone, Stout, and Tomaselli showed that the binary companion can induce transitions between bound and unbound states of a gravitational atom, effectively "ionizing" it, in a process analogous to the photoelectric effect in ordinary atoms. Ionization losses would not be just a small perturbation of the system: they can be larger than gravitational wave losses, and they contain sharp features that if detected would point directly to the quantum nature of the system \cite{2022PhRvD.105k5036B,2022PhRvL.128v1102B} (see Fig. \ref{fig:profile2}). 
The ionization energy losses, which can be interpreted as the backreaction on the orbit of dynamical friction, can be written as \cite{2022PhRvD.105k5036B}
\begin{equation}
    \dot{E}_\lab{ion} = \pm \frac{m_\lab{c}}{\mu}\sum_{\ell', m'} (m'-m) \Omega\, \big|\eta(\epsilon_*)\big|^2\, \Theta(\epsilon_*)\,.
    \label{eqn:ionization-power}
\end{equation}
where the sum runs over angular momentum states,  $\eta(\epsilon)=\langle\epsilon;\ell'm'|V|n\ell m\rangle$ is the level mixing induced by the gravitational perturbation $V$ of the companion, $\epsilon_*=-\mu\alpha^2/(2n^2)\pm(m'-m)\Omega$, and $\Theta$ is the Heaviside step function. The $\pm$ sign is for co-/counter-rotating orbits respectively \cite{2022PhRvD.105k5036B}. The analysis was generalised to inclined and eccentric orbits by Tomaselli, Spieksma and Bertone \cite{2023JCAP...07..070T,Tomaselli:2024bdd}.

\section{Gravitational Wave Probes of dark matter}
\label{sec:gw}

\begin{figure*}[t]
     \centering
         \includegraphics[width=0.8\textwidth]{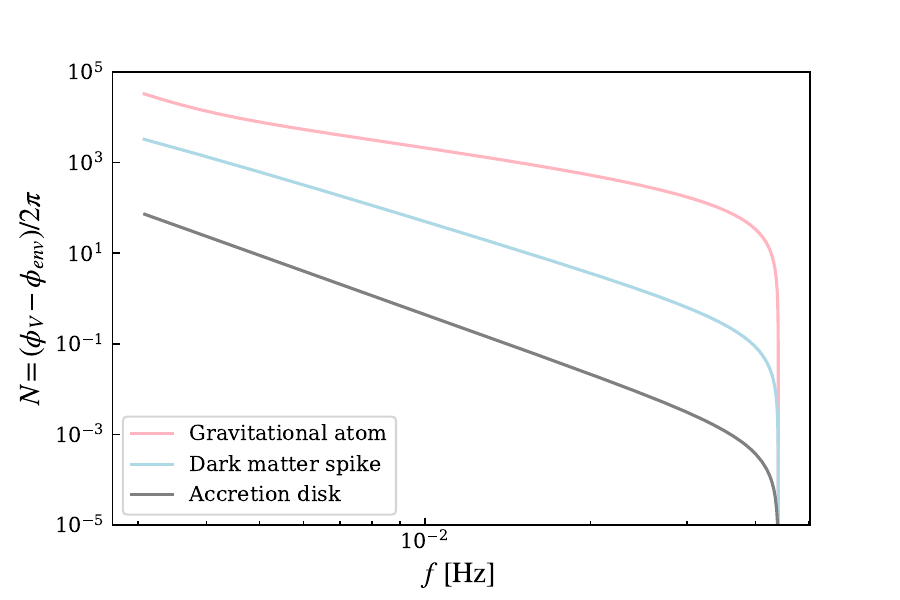}
        \caption{Dephasing with respect to a vacuum inspiral, for the environments in fig. 1. See Ref. \cite{2023NatAs...7..943C}, from which the figure is taken, for more details.}
        \label{fig:dephasing}
\end{figure*}

\textbf{Environment induced dephasing.} The rate of change of the separation of the binary is modified by the energy losses induced by the black hole environments:
\begin{equation}
    \dot{r} =  \frac{2r^2\dot{E}}{ G m_1 m_2}=\dot{r}_{\rm GW} + \dot{r}_{\rm env}.
\end{equation}
The frequency evolution and the phase of the signal can be calculated as:
\begin{equation}
\label{eq:fKep}
    f(t)=\frac{1}{\pi}\sqrt{\frac{G(m_1 + m_2)}{r(t)^3}}\,, \quad \Phi(f) = \int_f^{f_\mathrm{ISCO}} \frac{\mathrm{d}t}{\mathrm{d}f^\prime} f^\prime \,\mathrm{d}f^\prime\,,
\end{equation}
where $f_\mathrm{ISCO}$ is the frequency of the innermost stable circular orbit. The presence of dark matter modifies the time evolution of the separation and frequency, and manifests itself on the waveform as a dephasing with respect to a inspiral in vacuum. The \textit{dephasing}, defined as the difference in the number of cycles $N_{\rm cyc}$ from a given reference frequency until merger, between a given environment and the vacuum case is  :
\begin{equation} \label{eq:dephasing_definition}
    \delta\Phi = 2\pi (N_{\rm cyc, V} - N_{\rm cyc, env}) = \Phi_{\rm V} - \Phi_{\rm env}\,.
\end{equation}
The dephasing for different environments, with respect to a vacuum system with the same masses, is shown in Fig. \ref{fig:dephasing}. The phase shift enters into the calculation of the gravitational waveform via its second derivative with respect to time. For a review of enviromental effects, and their implications for gravitational wave measurements, see the review of Barausse, Cardoso, and Pani \cite{2014PhRvD..89j4059B}.

\textbf{Detectability, Discoverability, Measurability.} Coogan, Bertone, Gaggero, Kavanagh, and Nichols studied the prospects for detecting and characterizing dark matter overdensities around black holes through the analysis of the dephasing induced in the gravitational waveform of intermediate mass-ratio inspirals \cite{2022PhRvD.105d3009C}. They studied in particular the expected signal-to-noise ratio (which measures the \textit{detectability} of the inspiral with gravitational wave interferometers), and introduced a Bayesian framework to assess the prospects for discriminating dephased waveforms  against in-vacuum inspiral (\textit{discoverability}), and for measuring dark matter overdensity parameters (\textit{measurability}), assuming a detection with the future space interferometer LISA. Systems with chirp masses larger than $\mathcal{M} \sim \SI{16}{\solarmass}$ were found to be detectable out to $D_L \sim \SI{75}{\mega\parsec}$, and black holes surrounded by dark matter to be easily discriminated from GR-in-vacuum systems. In case of detection, the authors found that it would be possible to determine the halo's initial density normalization with a $\sim 15\%$ uncertaintainty, and the halo's slope with a $\lesssim 3\%$ uncertainty \cite{2022PhRvD.105d3009C}. Cole et al.  showed that dark matter overdensities around primordial black holes may be detectable with future ground based telescopes the Einstein Telescope or Cosmic Explorer. If detected, the properties of both the binary and of the dark matter distribution could be very precisely measured with one week's worth of data \cite{2023PhRvD.107h3006C}.

\textbf{Discs, Spikes, clouds.} The next generation of gravitational wave detectors will allow the exploration of the environments of intermediate mass ratio inspirals. Cole et al. showed that it is possible to accurately reconstruct the parameters of dark matter overdensities, accretion disks, and gravitational atoms, given the detection of a waveform with SNR of 15 and a duration of one year \cite{2023NatAs...7..943C}. By comparing the Bayesian evidence for the correct environmental template for a given signal, with that for an incorrect one, they found that waveforms contain enough information to confidently identify the correct environmental template. This implies that the risk of misinterpreting an environmental signal for a biased vacuum system, or the wrong type of environment, is extremely low. Furthermore, it is crucial to take environmental effects into account when searching for long-duration signals, as SNR losses are significant if the wrong template is used to fit the signal, and potential biases (e.g. in the chirp mass) can be large \cite{2023NatAs...7..943C}.

\textbf{Other probes of dark matter overdensities.} Although we focused here on dark matter annihilations and dephasing of the gravitational waveform, many other suggestions have been put forward to probe dark matter overdensities around black holes. One could for instance search for the modification of the black hole shadow induced by  dark matter \cite{2021ApJ...916..116N,2020PhRvD.102l4048E,2013A&A...554A..36L,2007PhRvD..75b3521P,2023JCAP...05..027C,2023arXiv230805544Y,2023arXiv230713553L}. Alternatively, it is possible to obtain interesting constraints from the analysis of stars observed around Sgr A* \cite{2020A&A...636L...5G,2007PhRvD..76f2001Z,2021ApJ...916..116N,2018A&A...619A..46L,2024MNRAS.527.3196S,2022NatSR..1215258C,2020arXiv200606219T,2023MNRAS.524.1075F,2023JCAP...11..063Z,2023arXiv231116228J,2023arXiv230809170L,Caiozzo:2024flz}.  

\textbf{Other results.} It is impossible in the space allowed by these proceedings to do justice to all the exciting work in this fast growing field of research. The interested reader will find more information on dynamical friction in dark matter overdensities around black holes in Refs. \cite{2023arXiv231118156A,2023arXiv231116539K,2023arXiv231107672B,2023arXiv230906498N,2023arXiv230905061C,2023arXiv230617158G,2023arXiv230617143S,2023arXiv230611787C,2023arXiv230517281D,2023PhRvD.107h3005K,2023ApJ...943L..11C,2022SCPMA..6500412L,2022PhRvD.106f4003D,2022PhRvD.106d4027S,2022arXiv220711145X,2022PhRvD.105l3018T,2022PhRvD.105f3029B,2020PhLB..81135944A,2020A&A...644A.147C,2020PhRvD.102f3022A,2019PhRvD.100d3013Y,2019ApJ...874...34Y,2018PhRvD..97f4003Y}. There are many other connections between dark matter and gravitational waves, that go beyond the environmental effects discussed here, including primordial black holes, exotic compact objects, direct dark matter detection with GW experiments, non-perturbative DM dynamics, and phase transitions. For a recent review, and many references, see the White Paper \textit{``Gravitational Wave Probes of Dark Matter''}, by Bertone et al. \cite{2019arXiv190710610B}. The broader implications for fundamental physics are reviewed in the White Paper \textit{`Black holes, gravitational waves and fundamental physics: a roadmap'} by Barack et al. \cite{2019CQGra..36n3001B}.  

\section{Conclusions.}

The interface between dark matter, black holes, and gravitational waves represents a new and exciting frontier in fundamental physics and astronomy. I provided in this article a brief overview of a few promising lines of research, but it is clear that we are standing at the edge of a vast expanse of uncharted territory. Much remains to be done to match the level of precision that will be required to interpret the next generation of gravitational wave interferometers, including: a full relativistic treatment of black hole environments and their joint evolution with the binary; a detailed study of degeneracies with GR effects; and the understanding of the interplay with transient orbital resonances and modified gravity. Furthermore, new data analysis strategies will be needed to meet the  challenge of analysing the long duration signals required by environmental studies, in view of the predicted background of thousands of galactic binaries, and of possible gaps in the data. Despite these difficulties, the challenge is worth undertaking: gravitational waves offer a unique window into the fabric of the universe, and might hold the key to unlock the longstanding mystery of dark matter.

\section*{Acknowledgements.} It is a pleasure to acknowledge the many collaborators who have contributed to the work discussed in this paper, including: Daniel Baumann, Pippa Cole, Adam Coogan, Djuna Croon, Darren Croton, Mattia Fornasa, Daniele Gaggero, Stefan Harfst, Theophanes Karydas, Bradley Kavanagh, David Merritt, Emmanuel Moulin, Priya Natarayan, David Nichols, Joe Silk, Thomas Spieksma, Guenter Sigl, John Stout, Marco Taoso, Gimmy Tomaselli, Marta Volonteri, Renske Wierda, Naoki Yoshida, and Andrew Zentner. I gratefully acknowledge the Department of Physics of Columbia U. and the Italian Academy for Advanced Studies in America, where part of this work was carried out.

%% If you have bibdatabase file and want bibtex to generate the
%% bibitems, please use
%%
 \bibliographystyle{elsarticle-num} 
 \bibliography{GW}

%% else use the following coding to input the bibitems directly in the
%% TeX file.

% \begin{thebibliography}{00}

% %% \bibitem{label}
% %% Text of bibliographic item

% \bibitem{}

% \end{thebibliography}
\end{document}